\documentstyle[sprocl]{article}

\def\kv{{\bf k}}
\def\g{{\gamma}}

\begin{document}

\title{LATTICE COLOR GROUPS OF QUASICRYSTALS}

\author{RON LIFSHITZ}

\address{Condensed Matter Physics 114-36, California Institute of
Technology,\\ Pasadena, CA 91125, U.S.A.\ \ (E-mail:
lifshitz@cco.caltech.edu)}

\maketitle

\abstracts{Lattice color groups are introduced and used to study the
partitioning of a periodically- or quasiperiodically-ordered set of
points into $n$ symmetry-related subsets.  Applications range from
magnetic structure to superlattice ordering in periodic and
quasiperiodic crystals.}

\section{Introduction}

It is well known that the points of a square lattice may be
partitioned into two square subsets, related by a translation, but
that the same is not true for triangular lattices. This allows a
certain kind of anti-ferromagnetic order in tetragonal crystals which
is not possible for hexagonal crystals. One often encounters similar
situations in which a set of lattice points is to be partitioned into
$n$ ``symmetry-related'' subsets (to be properly defined below). If
the lattice points correspond, for example, to atomic positions in a
crystal then the different subsets may correspond to different
chemical species or to $n$ different orientations of a magnetic
moment.  One may also single out just one of the subsets to play a
significant role such as in describing superlattice ordering.  We
shall address here the generalization of this question to
quasiperiodic crystals. In doing so we shall introduce some aspects of
the theory of color symmetry for periodic and quasiperiodic
crystals.\cite{rmp} Please consult Ref.~1 for complete detail and a
rigorous derivation of the results given here.

\section{Color symmetry}

By associating one of $n$ distinct colors with each site of a crystal
one produces a {\it colored crystal\/} and a partitioning, as
described above. The colored crystal is said to have {\it color
symmetry\/} if rotations (and, in the special case of periodic
crystals, translations) that are symmetry operations of the uncolored
crystal may be combined with global permutations of the $n$ colors to
become symmetry operations of the colored crystal.  

What do we mean when we say that a rotation is a symmetry of a
quasiperiodic crystal?  In the case of periodic crystals we mean that
the rotation leaves the crystal {\it invariant\/} to within a
translation, but this is in general not the case for quasicrystals.
The key to redefining symmetry for quasiperiodic crystals is the
notion of indistinguishability.  Two crystals are {\it
indistinguishable\/} if they contain the same spatial distribution of
bounded substructures of arbitrary size. For a (proper or improper)
rotation to be in the {\it point group\/} $G$ of a quasiperiodic
crystal it is only required to leave it indistinguishable and not
necessarily invariant. It can be shown that if two {\it periodic\/}
crystals are indistinguishable then they differ at most by a
translation, thus reducing the new definition of point group back to
the traditional one. A more detailed account of these concepts and
their use in generalizing crystallography to treat quasiperiodic
crystals may be found elsewhere.\cite{physa}

The point group $G$ of a colored quasiperiodic crystal is subsequently
defined as the set of rotations which leave it indistinguishable to
within a permutation of the colors. The {\it color point group\/} of
the crystal consists of all pairs $(g,\g)$ leaving it
indistinguishable, where the $g$'s are elements of the point group $G$
and the $\g$'s are color permutations. In general, there can be many
different $\g$'s associated with a single point group rotation $g$. Of
particular interest are the color permutations which may be paired
with the identity rotation to leave the colored crystal
indistinguishable.  These color permutations form the {\it lattice
color group\/} which is the focus of our discussion here. In the
special case of periodic crystals these are permutations which when
combined with a translation leave the crystal invariant.

\begin{figure}[p]
\begin{center}
\begin{picture}(440,440)
\put(-110,-90){\special{hscale=90 vscale=90 psfile=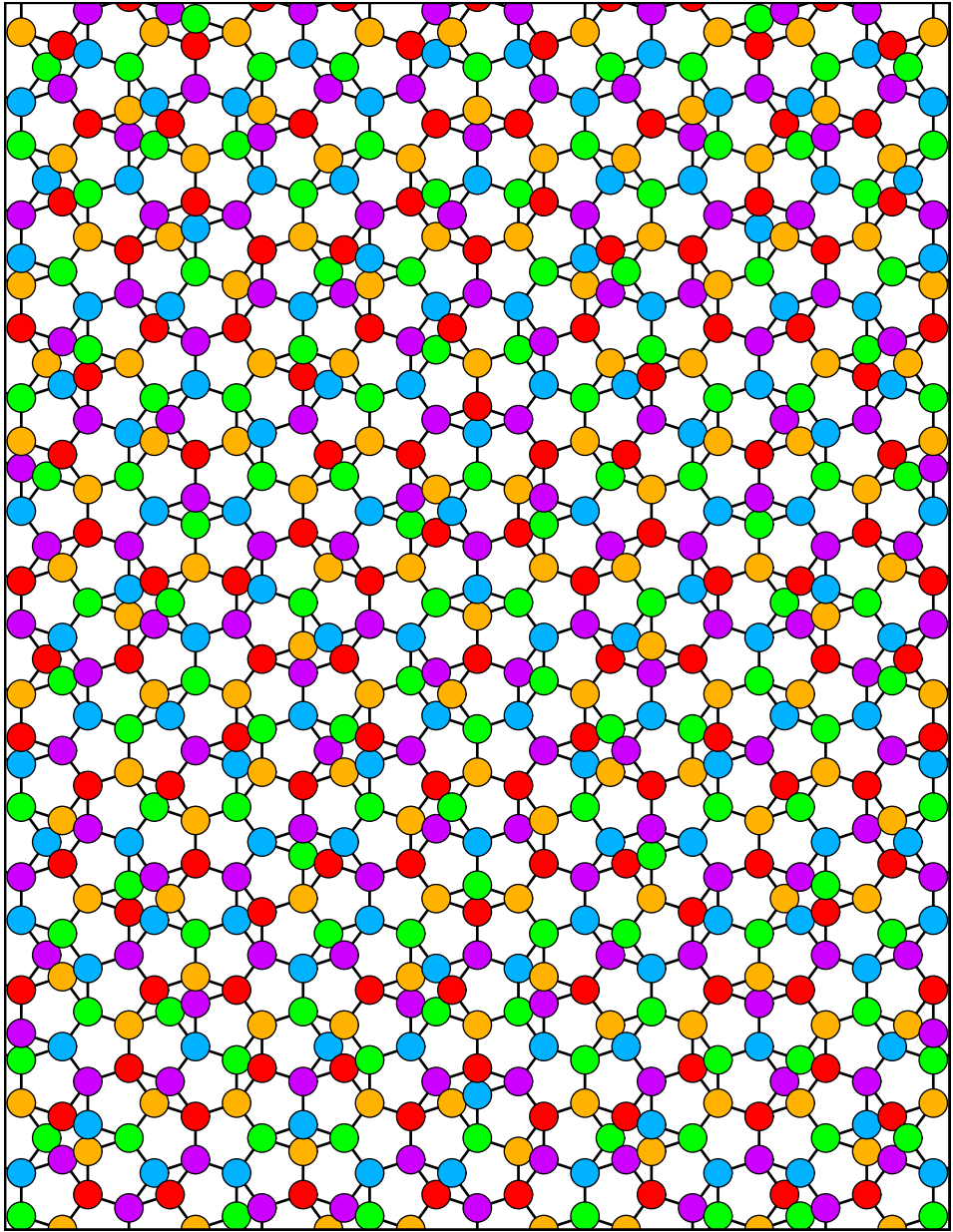}}
\end{picture}
\end{center}
\caption{A symmetric partitioning with 5 colors of the vertices of a
10-fold Penrose tiling, originally introduced by
L\"uck.\protect\cite{luck} The lattice color group is the cyclic group
of order~5, generated by the permutation (red, purple, blue, green,
orange).}
\label{fig.pen5}
\end{figure}


\def\tt{\thinspace\thinspace}

\begin{table}[t]
\caption{Indices of invariant sublattices of standard 2-dimensional
$N$-fold lattices, giving the possible numbers of colors in a
symmetric partitioning of a 2-dimensional $N$-fold
quasicrystal. The sublattices are only required to be invariant under the
point group $G=n$ and not under the full point group of the $N$-fold
lattice which also contains mirror reflections.  }
\label{tab.indices}

\smallskip
{\scriptsize
\centerline{
\setbox\strutbox=\hbox{\vrule height7pt depth3pt width0pt}
\vbox{\offinterlineskip
\hrule
\hrule
\halign{\strut \vrule\vrule\hfil\tt $#$\tt\hfil
              &\vrule\hfil\tt $#$\tt\hfil 
              &\vrule\tt $#$\tt\tt\hfil\vrule\vrule\cr
$Lattice$ & $Point Group$ & \hfil $Indices of sublattices,$ \cr 
$order $ N & G & \hfil $invariant under $ G=n\cr
\noalign{\hrule}
\noalign{\hrule} 
\multispan3{\strut\vrule\vrule\hfil\tt  Rank
 2\tt\hfil\vrule\vrule}\cr
\noalign{\hrule}
\noalign{\hrule}
4 & 4 & 1, 2, 4, 5, 8, 9, 10, 13, 16, 17, 18, 20, 25, 26, 29, 32, 34, 36,
37, \ldots
\cr
\noalign{\hrule}
6 & 3, 6 & 1, 3, 4, 7, 9, 12, 13, 16, 19, 21, 25, 27, 28, 31, 36, 37,
39, 43, \ldots \cr
\noalign{\hrule}
\noalign{\hrule}
\multispan3{\strut\vrule\vrule\hfil\tt  Rank
 4\tt\hfil\vrule\vrule}\cr
\noalign{\hrule}
\noalign{\hrule}
8 & 8 & 1, 2, 4, 8, 9, 16, 17, 18, 25, 32, 34, 36, 41, 49, 50, 64, 68,
72, \ldots \cr
\noalign{\hrule}
10 & 5, 10 & 1, 5, 11, 16, 25, 31, 41, 55, 61, 71, 80, 81, 101, 121,
125, 131, \ldots \cr
\noalign{\hrule}
12 & 12 & 1, 4, 9, 13, 16, 25, 36, 37, 49, 52, 61, 64, 73, 81, 97,
100, 109, \ldots \cr
\noalign{\hrule}
\noalign{\hrule}
\multispan3{\strut\vrule\vrule\hfil\tt  Rank
 6\tt\hfil\vrule\vrule}\cr
\noalign{\hrule}
\noalign{\hrule}
14 & 7, 14 & 1, 7, 8, 29, 43, 49, 56, 64, 71, 113, 127, 169, 197, 203,
211, 232, \ldots \cr
\noalign{\hrule}
18 & 9, 18 & 1, 3, 9, 19, 27, 37, 57, 64, 73, 81, 109, 111, 127, 163,
171, 181, \ldots \cr
\noalign{\hrule}
\noalign{\hrule}
\multispan3{\strut\vrule\vrule\hfil\tt  Rank
 8\tt\hfil\vrule\vrule}\cr
\noalign{\hrule}
\noalign{\hrule}
16 & 16 & 1, 2, 4, 8, 16, 17, 32, 34, 49, 64, 68, 81, 97, 113, 128,
136, 162, \ldots \cr
\noalign{\hrule}
20 & 20 & 1, 5, 16, 25, 41, 61, 80, 81, 101, 121, 125, 181, 205, 241,
256, \ldots \cr
\noalign{\hrule}
24 & 24 & 1, 4, 9, 16, 25, 36, 49, 64, 73, 81, 97, 100, 121, 144, 169,
193, \ldots \cr
\noalign{\hrule}
30 & 15, 30 & 1, 16, 25, 31, 61, 81, 121, 151, 181, 211, 241, 256,
271, 331, \ldots \cr
\noalign{\hrule}
\noalign{\hrule}}}}
}
\end{table}

\section{Lattice Color Groups and Invariant Sublattices}

For the purpose of our discussion here we define a {\it symmetric
partitioning\/} of a quasiperiodically-ordered set of points to be an
$n$-coloring of the set, satisfying the following requirements: ({\it
a}) The point group $G$ of the colored set is the same as that of the
uncolored one; and ({\it b}) the lattice color group is transitive on
the $n$ colors. To elaborate, we require that any rotation in the
point group of the uncolored set of points may be combined with a
color permutation to leave the colored set indistinguishable, and that
for any pair of differently-colored subsets there exists at least one
permutation in the lattice color group, taking one into the other. It
follows from ({\it a}) and ({\it b}) that each individually-colored
subset is also left indistinguishable by the elements of $G$.

Colored periodic crystals, satisfying the requirements above, have
been studied previously.\cite{harker} To each element of the crystal's
periodic lattice $T$ of translations corresponds a unique color
permutation. One finds that the sublattice $T_0$, associated with the
identity color permutation, is invariant under the point group $G$ of
the crystal; its index in $T$ is equal to the number of colors $n$;
and the quotient group $T/T_0$ is isomorphic to the lattice color
group. The classification of lattice color groups, compatible with
crystals with a given lattice $T$ and point group $G$, thus reduces to
the classification of sublattices $T_0$ of $T$ that are invariant
under $G$.

In the case of quasiperiodic crystals one has no lattice of
translations but as it turns out,\cite{rmp} a similar situation arises
in Fourier space which holds both for periodic and quasiperiodic
crystals. The (Fourier) lattice $L_0$ (called by some ``the Fourier
module''), consisting of all the wave vectors appearing in the
diffraction diagram of the uncolored crystal, is a sublattice of the
(Fourier) lattice $L$, consisting of all the wave vectors that would
appear in a diffraction diagram due to the sites of a single color in
the colored crystal. This is a phenomenon, familiar from structures
with superlattice ordering, where additional weak satellites appear in
the diffraction diagram turning the original lattice $L_0$ into $L$.
Such diffraction diagrams, exhibiting superstructure ordering, have
been observed in decagonal AlCoNi quasicrystals~\cite{deca} as well as
in icosahedral AlPdMn quasicrystals.\cite{ico}

One finds that the sublattice $L_0$ is invariant under the point group
$G$ of the full lattice $L$; the index of $L_0$ in $L$ is equal to the
number of colors; and the quotient group $L/L_0$ is isomorphic to the
lattice color group.  Again, the classification of lattice color
groups is reduced to the classification of invariant (Fourier)
sublattices.


Consider as a first example the partitioning of a standard
2-dimensional $N$-fold quasicrystal ($N<46$) into {\it two\/}
indistinguishable parts, allowing the type of anti-ferromag\-netic
order discussed in the introduction. The lattice $L$ of a standard
$N$-fold quasicrystal may be generated by an $N$-fold star of wave
vectors of equal length.  For the required partitioning, $L$ must
contain an invariant sublattice $L_0$ of index 2.  One can show that any sum,
containing an {\it odd\/} number of vectors from the generating star,
must not belong to $L_0$. It then follows that invariant sublattices of
index~2 can exist only if $N$ is a power of~2. Niizeki \cite{niizeki}
has shown such a 2-coloring of the vertices of an octagonal tiling.

Consider finally the possible values of $n$, compatible with a
symmetric partitioning of a given standard 2-dimensional $N$-fold
quasicrystal ($N<46$). Because all $N$-fold lattices with $N<46$ are
standard any invariant sublattice must itself be a standard $N$-fold
lattice. Any arbitrary vector $\kv\in L$ belongs to an $N$-fold star
which generates a sublattice $L_0$ of $L$. One can thus generate all
the sublattices of $L$ simply by letting $\kv$ run through all wave
vectors in $L$.  The index of the sublattice is just the magnitude of
the determinant of the matrix which gives the basis of $L_0$ in terms
of the basis of $L$. Results for lattices up to rank~8, taken from
Ref.~1, are reproduced in Table~1. An example of a 5-coloring of the
vertices of a decagonal Penrose tiling, first introduced by
L\"uck,\cite{luck} is shown in Figure~1.  A different approach for
calculating invariant sublattices, which involves generating functions
of Dirichlet series, is used by Baake {\it et al.}\cite{baake} and
yields the same results.

\smallskip
This work is supported by the California Institute of Technology
through its Prize Fellowship in Theoretical Physics.

\end{document}